# Doping-induced spin Hall ratio enhancement in A15-phase, Ta-doped β-W thin films

Mohsin Z. Minhas[1], Avanindra K. Pandeya[1], Bharat Grover[1], Alessandro Fumarola[1], Ilya Kostanovskiy[1], Wolfgang Hoppe[2], Georg Woltersdorf[2], Amilcar Bedoya-Pinto[1*], Stuart S. P. Parkin[1] and Mazhar N. Ali[1*]

[1]Max Planck Institute for Microstructure Physics, 06120 Halle (Saale), Germany

[2]Institute of Physics, Martin Luther University Halle-Wittenberg, Von-Danckelmann-Platz 3, 06120 Halle (Saale), Germany

**Abstract:**

As spintronic devices become more and more prevalent, the desire to find Pt-free materials with large spin Hall effects is increasing. Previously it was shown that β-W, the metastable A15 structured variant of pure W, has charge-spin conversion efficiencies on par with Pt, and it was predicted that β-W/Ta alloys should be even more efficient. Here we demonstrate the enhancement of the spin Hall ratio (SHR) in A15-phase β-W films doped with Ta ($W_{4-x}Ta_x$ where $0.28 <= x <= 0.4$) deposited at room temperature using DC magnetron co-sputtering. In close agreement with theoretical predictions, we find that the SHR of the doped films was ~9% larger than pure β-W films. We also found that the SHR's in devices with $Co_2Fe_6B_2$ were nearly twice as large as the SHR's in devices with $Co_4Fe_4B_2$. This work shows that by optimizing deposition parameters and substrates, the fabrication of the optimum $W_3Ta$ alloy should be feasible, opening the door to commercially viable, Pt-free, spintronic devices.

**Introduction/Background:**

Recently, the spin Hall effect (SHE) has received a great deal of attention not only from a fundamental physics perspective but also for technological applications in the area of memories, logic and sensors[1][2]. The SHE is the conversion of a longitudinal current density into a transverse spin current density, which is characterized by the spin Hall ratio (SHR), $\theta_{SH}=J_S/J_C$ and materials with large SHRs are desired for spintronic applications. Note that SHR and SHA (spin Hall angle) have been used interchangeably in literature, however the SHA is rigorously defined as the $\tan^{-1}(J_S/J_C)$. Previous studies have demonstrated only a handful of materials that can be sputter-deposited and which exhibit large SHRs at room temperature: these include Pt/doped Pt [3], β-Ta [4], β-W [5] and oxygen doped β-W [6]. There is a technological desire towards Pt-free spintronics and among these candidates, β-W and oxygen doped β-W, have the highest resistance (resulting in larger power requirements) but also show the largest SHR, which is approximately -0.35 and -0.5, respectively [5][6][7] [8]. β-W is a metastable phase of W that exhibits the cubic A15 structure (SG# 223, Pm-3n) which has two distinct crystallographic sites with two atoms located on each face of the cubic cell. The α-W phase, on the other hand, is the most stable form of W that crystallizes in body centered cubic structure, while exhibiting a smaller SHE [9][10][11][12].

There are two categories of mechanisms which govern the SHE: extrinsic and intrinsic. The former arises from spinful electron scattering from impurities or defects typically due to spin-orbit coupling (SOC), while in the latter case the electron is deflected during its transport in-between scattering events [13][14][15]. This occurs due to presence of spin Berry curvature (SBC) generated by, for example, SOC-gapped Dirac points in the material's electronic structure [16]. Materials with large intrinsic SHEs are becoming popular as the spin current generating layer of spintronic devices because recently, using concepts form topological physics, a series of sputter-able, low-cost, and giant SHE alloys have been predicted [16].

Among these, $W_{4-x}Ta_x$ at x=1 was predicted to host the highest spin hall conductivity (SHC); approximately 20% more than pure β-W[16][17].

Here we experimentally demonstrate that $W_{4-x}Ta_x$, where 0.28 <= x <= 0.4 can be stabilized in the A15 structure on sapphire substrates via room temperature co-sputtering. Using spin-torque ferromagnetic resonance (ST-FMR), we further show that the $W_{4-x}Ta_x$ films host higher SHRs than their doped counter parts (~9%), in accordance with theoretical calculations [16][17]. In addition, we found a significant dependence on the type of CoFeB (4:4:2 vs 2:6:2) magnetic layer used; a 2x fold enhancement in the measured SHR for both the pure W and the Ta-doped films when using the 2:6:2 CoFeB composition (with respect to 4:4:2). Future studies on these $W_3Ta$ alloys on low-cost $SiO_2$ substrates will open the door to facile device fabrication of Pt-free spintronic devices.

**Methods, Results and Discussion:**

We deposited the pure β-W and $W_{4-x}Ta_x$ film stacks by DC magnetron co-sputtering on $Al_2O_3$ substrates at room temperature. The base pressure of the chamber before the deposition was ~$10^{-9}$ Torr: the Ar sputtering pressure during deposition was 3 mTorr. Prior to deposition, the substrates were cleaned by a standard multi-step wet cleaning method [18] with an additional annealing step (annealed at 1200°C for 4 hours in atmospheric conditions). In order to get homogeneous film thickness, the substrate holder was rotated at 10 rpm. Pure β-W and $W_{4-x}Ta_x$ films were deposited using sputtering powers of 160 watt (W) and 10 watt (Ta), yielding deposition rates of 5.6 nm/min for the β-W films and 6 nm/min for the $W_{4-x}Ta_x$ films. Two different compositions of magnetic layers (i.e $Co_4Fe_4B_2$ and $Co_2Fe_6B_2$) were used while keeping their thickness constant at 3 nm. Highly resistive TaN (2nm) was deposited as a capping layer to prevent oxidation of the active layers. The complete stacks for SHR measurements were W(15)/$Co_4Fe_4B_2$(3)/TaN(2) (denoted as S1) and $W_{4-x}Ta_x$(15)/$Co_4Fe_4B_2$(3)/TaN(2) (denoted as S2), and also W(15)/$Co_2Fe_6B_2$(3)/TaN(2) (denoted as S3) and $W_{4-x}Ta_x$(15)/$Co_2Fe_6B_2$(3)/TaN(2) (denoted as S4), where the number in parentheses is the film thickness in nm of that particular layer.

X-ray diffraction was used to confirm the A15 phase of the film. The diffraction peaks, (200) and (211) indicate that the β-W phase was successfully grown in the A15 crystal structure as shown in Fig.1(a). Note that no extra peaks indicating Ta ordering and supercell formation, are visible, implying Ta/W solid solution formation. X-ray photoelectron spectroscopy (XPS) chemical identification and depth profiling was used to determine the composition of $W_{4-x}Ta_x$ films as shown in Fig.1(b), where the W peaks match extremely well with bulk metallic W while the Ta peaks are slightly upshifted from bulk metallic Ta. The Ta doping is between 7-10% (0.28 <= x <= 0.4). Ideally, $W_3Ta$ should be fabricable, however we were unable to achieve such high concentration of Ta while maintaining the A15 structure. Higher temperature as well as lower power deposition were found to promote the α-phase; using different substrates may be able to stabilize higher Ta-doping concentrations in the A15 phase. The elemental composition of the $W_{4-x}Ta_x$ films was tuned by sputtering power of W and Ta targets. The effective magnetization $M_{eff}$ is extracted by using the Kittel relation [19] as shown in Fig.1(c). The saturation magnetization ($M_s$), measured by vibrating sample magnetometry (VSM), and the resistivity (four-point probe method) was 893 emu/cm$^3$ and $\rho \approx 420$ μΩ cm, respectively for $Co_4Fe_4B_2$. For the $Co_2Fe_6B_2$ samples, these values were 970 emu/cm$^3$ and $\rho \approx 272$ μΩ cm. The resistivity of the W films was $\rho \approx 220$ μΩ cm and $W_{4-x}Ta_x$ was in the range of $\rho \approx$ 235-310 μΩ cm.

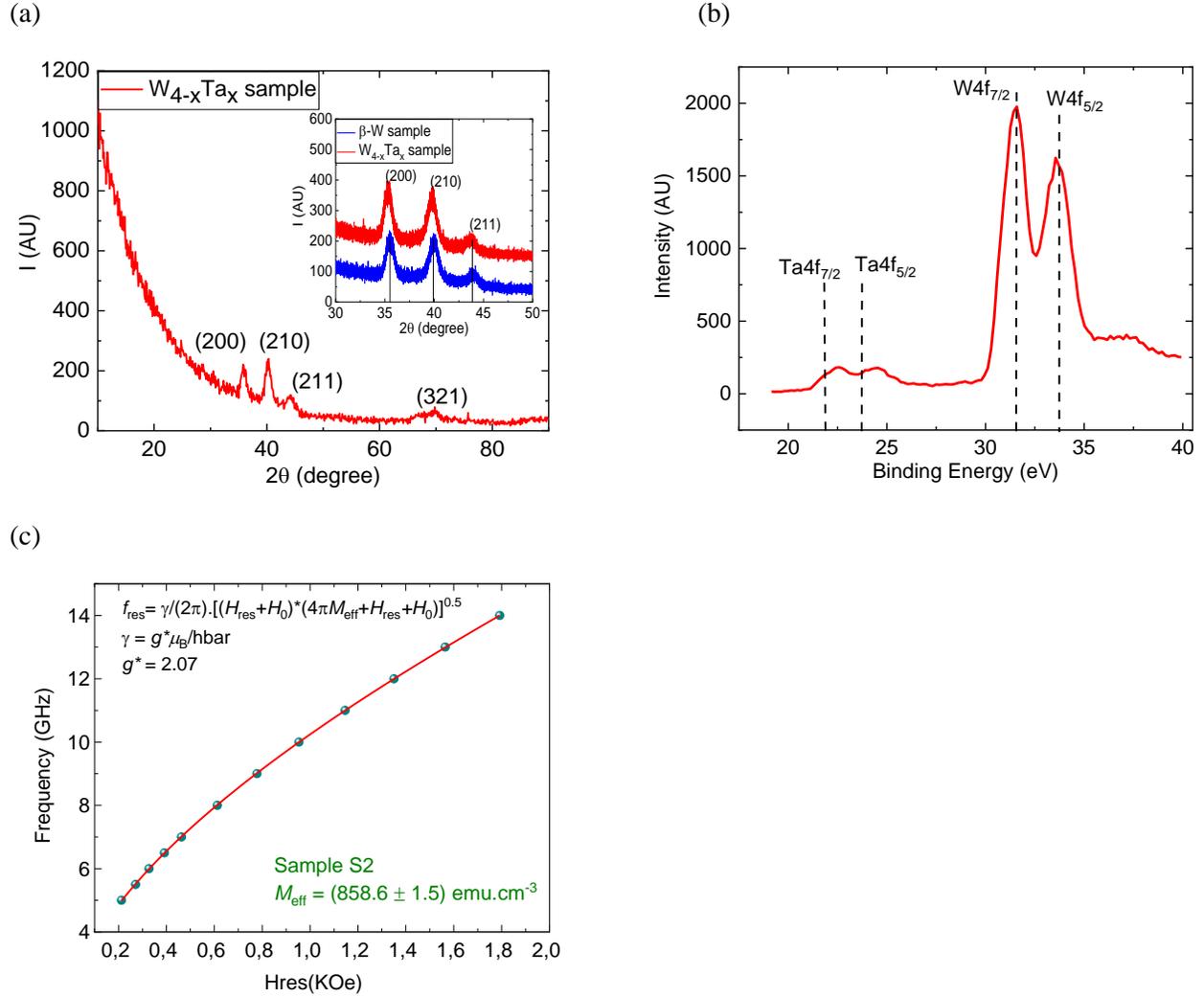

Figure 1. (a) X-ray diffraction patterns for $W_{4-x}Ta_x$ thin film deposited at room temperature, Bragg peaks are labelled with corresponding diffraction planes. Inset: show the β-W and $W_{4-x}Ta_x$ thin films diffraction patterns in short angle range near the peaks. (b) XPS spectra shows the atomic composition of $W_{4-x}Ta_x$ thin film, black dotted lines indicate the peak positions for the bulk W and Ta metals. (c) Resonance frequency as a function of bias magnetic field, the effective magnetization $M_{eff}$ is extracted by using the Kittel relation [19].

The multilayer stacks were patterned to microstrips with dimensions specially designed for spin-torque resonance experiments (schematic and actual microdevice shown in Figures 2a and b). We employ the spin-torque ferromagnetic resonance (ST-FMR) method to determine the spin Hall ratio.

In typical ST-FMR measurements of bilayer systems, several process occur: 1) by the application of a radiofrequency (rf) charge current along the film plane, an oscillating transverse spin current is generated in the non-magnetic ((β-W or $W_{4-x}Ta_x$) layer via the inverse spin hall effect (ISHE). 2) This spin current diffuses along z-direction and, if it is effectively injected into the adjacent magnetic layer ($Co_4Fe_4B_2$ or $Co_2Fe_6B_2$), it will exert an oscillating spin torque (ST) that influence the magnetization dynamics of the magnetic layer. The magnitude of the spin-torques will thus effect parameters such as lineshape and/or linewidth of the ferromagnetic resonance. 3) A mixing DC voltage, $V_{mix}$ arises from the oscillating anisotropic magnetoresistance, spin hall magnetoresistance, and the applied RF current [3][6][20][21], and is thus used to visualize the changes in the ferromagnetic resonance induced by spin-torques. Practically, this is done by measuring $V_{mix}$ as a function of magnetic field and/or frequency. We applied the magnetic field at 45° (or 135°) to a 10 μm × 10 μm microstrip of the stacks with the DC and RF currents applied though an RF probe and consequently measured the DC response across a bias Tee. The output voltage was then fitted by using equation (1).

$$V_{mix} = V_0 + V_{sym} \frac{\Delta^2}{\Delta^2 + (H_{ext} - H_0)^2} + V_{as} \frac{\Delta(H_{ext} - H_0)}{\Delta^2 + (H_{ext} - H_0)^2} \quad (1)$$

Here $V_0$ is the DC offset, $V_{sym}$ is the coefficient of the symmetric part of the Lorentzian, $V_{as}$ is the coefficient of anti-symmetric part of the Lorentzian, $\Delta$ is linewidth, $H_{ext}$ is the applied external field and $H_0$ is resonance field.

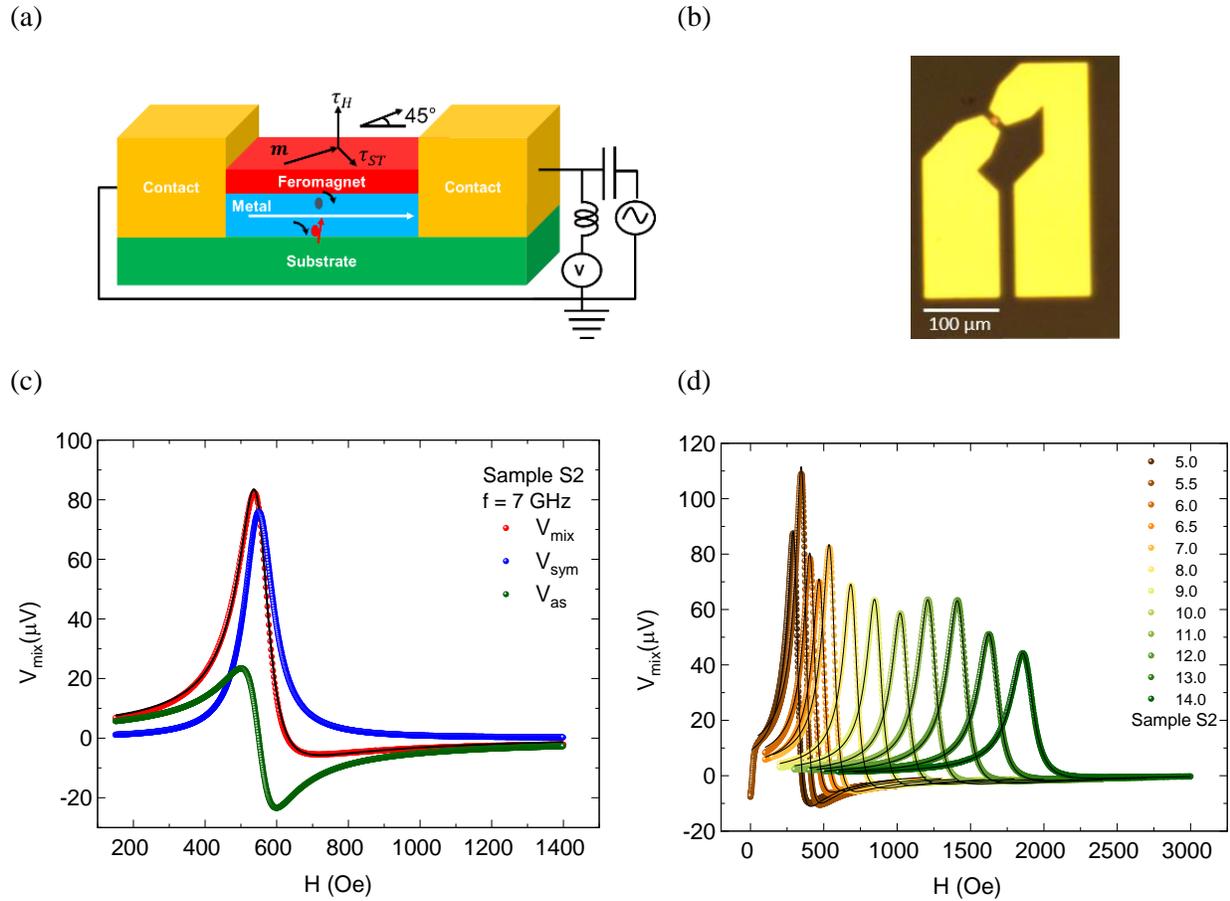

Figure 2. (a) Schematic illustration of a typical ST-FMR measurement arrangement in a bilayer system. (b) Microscopic image of device used for ST-FMR measurements. (c) Typical ST-FMR signal (red dots) of a 10 μm × 10 μm $W_{4-x}Ta_x$/$Co_4Fe_4B_2$ device at 7 GHz, fitted to Eq. 1 (solid black line) with symmetric Lorentzian ($V_{sym}$ blue curve) and antisymmetric Lorentzian ($V_{as}$ green curve) components. (d) ST-FMR spectra measured for 5-14 GHz at zero DC bias.

Figure 2(c) shows the STFMR signal (red dots are measurement data from $W_{4-x}Ta_x$/$Co_4Fe_4B_2$) at 7 GHz with the symmetric Lorentzian ($V_{sym}$) and antisymmetric Lorentzian ($V_{as}$) components, respectively

(extracted from fits to equation 1). The predominantly symmetric component suggests a high damping-like torque in the magnetic layer [3]. A large amplitude (50 to 100μV) of the ST-FMR voltage highlights the high quality of the magnetic sensing layer (CoFeB) and enables both a lineshape and linewidth analysis [6] for the calculation of the spin-torque efficiency. We rely here on the linewidth analysis, which directly reflects the amount of generated spin-torque (linewidth broadening) as the charge current is increased through the bilayer. Hence, the FMR linewidth as a function of DC current $I_{dc}$ for the all samples was fitted at frequencies ranging from 10-13 GHz as shown in Figure 3a and b, and the slope of these plots ($\delta\Delta H/\delta I_{DC}$) is used to determine the SHR using equation (2).

$$\theta_{sc} = \frac{\delta\Delta H/\delta I_{DC}}{\frac{2\pi f}{\gamma}\frac{sin\phi}{(H_0 + 0.5M_{eff})\mu_0 M_s t}\frac{\hbar}{2e}}\frac{R_{FM} + R_{metal}}{R_{FM}}A_c \qquad (2)$$

Here $\theta_{sc}$ is SHR (spin-to-charge conversion efficiency), $\delta\Delta H/\delta I_{DC}$ is the change in linewidth per unit applied current, $f$ is the applied frequency, $\phi$ is the angle between the current and the applied magnetic field, $M_{eff}$ is the effective magnetization, $M_s$ is the saturation magnetization, $t$ is the thickness of the magnetic layer, $R_{FM}$ is the resistance of the ferromagnetic layer, $R_{metal}$ is the resistance of the metallic layer and $A_c$ is the cross sectional area of the device. The effective magnetization (as shown in Fig.1(c)) and Gilbert damping in the magnetic layer is extracted from the Kittel dispersion fit of the ST-FMR signal at different frequencies as shown in Figure 2 (d) and outlined in Table 1 for the various film compositions.

(a) 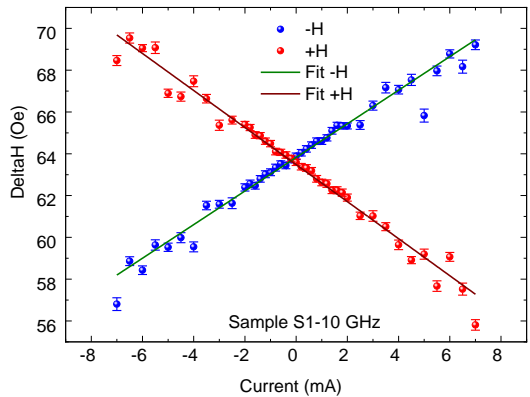

(b) 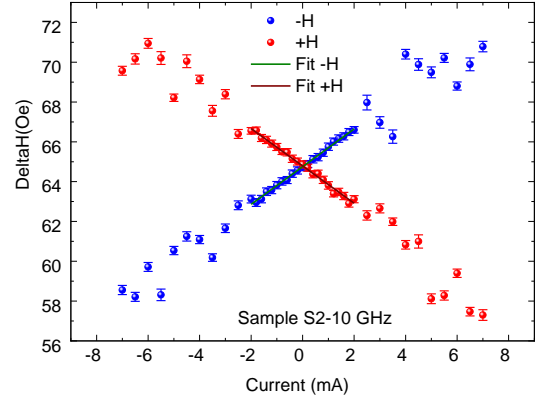

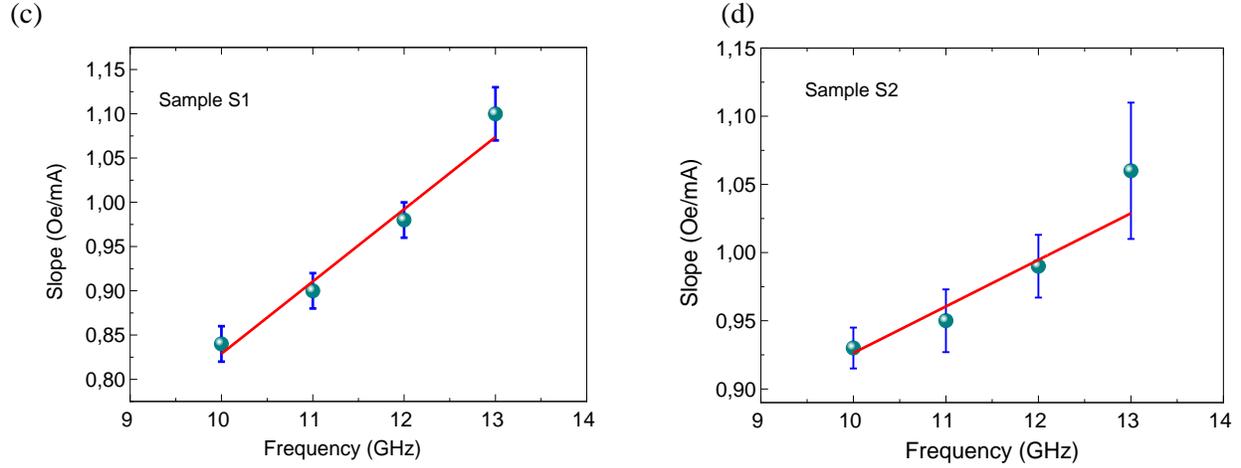

Figure. 3. (a and b) The FMR linewidth as a function of DC current $I_{dc}$ for the W/Co$_4$Fe$_4$B$_2$ and W$_{4-x}$Ta$_x$/Co$_4$Fe$_4$B$_2$ bilayers device respectively at f=10 GHz. (c and d) Shows the slope as a function of frequency of the applied rf current in the bilayer device, the linear scaling is in accordance with equation (2).

Table 1: The spin Hall ratio, Gilbert damping coefficient and effective magnetization extracted by STFMR measurements for all the samples.

| Sample | FM/NM layers | Gilbert damping coefficient α | Effective magnetization $M_{eff}$ (emu.cm$^{-3}$) | Spin hall ratio |
|---|---|---|---|---|
| S1 | W/Co$_4$Fe$_4$B$_2$ | 0.014 ± 0.00113 | 874 ± 1.4 | 0.0118 ± 0.002 |
| S2 | W$_{4-x}$Ta$_x$/Co$_4$Fe$_4$B$_2$ | 0.014 ± 0.00113 | 858.6 ± 1.5 | 0.0129 ± 0.0013 |
| S3 | W/Co$_2$Fe$_6$B$_2$ | 0.012 ± 0.00113 | 1019.6 ± 1.2 | 0.0225 ± 0.0003 |
| S4 | W$_{4-x}$Ta$_x$/Co$_2$Fe$_6$B$_2$ | 0.014 ± 0.0004 | 854 ± 1.5 | 0.0240 ± 0.0004 |

From analyzing the slopes (Figure 3c and 3d) and equation 2, the SHR of sample S1 is found to be 0.0118 ± 0.002 and for sample S2 is found to be 0.0129 ± 0.0013 which is approximately 9 % larger than sample S1. Similarly the SHR obtained for sample S3 is 0.0225 ± 0.0003 and for sample S4 is 0.024 ± 0.0004 which is approximately 7 % larger than sample S3 as shown in Table 1(supplementary Figure 1). Our observations are in agreement with previous predictions that A15 phase W$_{4-x}$Ta$_x$, where $0.28 < x < 0.4$, would result in ~10% increase in the intrinsic SHC and, correspondingly, SHR (due to a negligible change in longitudinal resistivity from Ta doping) [16][17].

We assume that the source of the SHE in our films is dominantly intrinsic like in pure β-W (due to band structure of the material) rather than extrinsic. This expectation comes from 1.) the small amount of doping, 2.) the fact that Ta and W are highly miscible and result in W/Ta solid solutions, as implied by XRD and 3.) that W and Ta are extremely close in SOC strength. The SOC argument is critical in distinguishing the effects of light dopants (e.g. oxygen or nitrogen) and heavy ones (such as Ta) in β-W films; since extrinsic spin Hall mechanisms depend on defects/dopants causing sudden fluctuations in the local spin-orbit field as an electron travels through the lattice [15], the extrinsic effect of Ta/W substitutional defects is expected to be small compared with O/W doping.

The lower values of SHR obtained in this study compared to previous studies on pure β-W [5] may be explained by interfacial transparency [22]. During the transmission of spin from the nonmagnetic layer to the ferromagnetic layer, spin loss can occur either by spin backflow or spin memory loss due to spin scattering at the interface [23][24][25]. The film microstructure at the interface can have additional impact on these interfacial effects and the previously mentioned studies were using $SiO_2$ substrates rather than sapphire, as we have. Contribution of these kinds of effects would ultimately underestimate the SHC and SHR values of the spin hall material. For example, it has been shown that the apparent difference in spin-orbit torque efficiency in Pt/Py films as compared with Pt/Co films can be explained by the spin transparency difference of the interfaces [22]. By factoring in the interface transparency between the ferromagnetic and nonmagnetic layers, the intrinsic SOT efficiency in Pt was determined to be ~0.3 [23] as opposed to the previously measured value of ~0.06 [3][26].

The enhanced value of the SHR for both pure β-W and Ta-doped films when using the Fe-rich $Co_2Fe_6B_2$ may be ascribed to an interfacial effect; Co-rich and Fe-rich CoFeB layers may have different interfacial energies at the interface adjacent to the nonmagnetic layer as well as TaN capping layer, which contributes to formation of a thin, 'dead' magnetic layer at the interfaces [27–29]. These magnetic dead layers are usually formed by intermixing at the interfaces. Generally speaking, the Fe-rich $Co_2Fe_6B_2$ is considered a better choice of magnetic layer because of its better perpendicular magnetic anisotropy (PMA) and tunneling magnetoresistance (TMR); both are important characteristics for spintronic devices [30,31]. Cross-sectional TEM comparison of β-W films grown using identical deposition conditions on both $SiO_2$ and sapphire substrates with both types of CoFeB layers may be useful in fully understanding these effects.

**Conclusion:**

In summary, we have demonstrated the enhancement of the spin Hall ratio in A15-phase β-W films doped with Ta ($W_{4-x}Ta_x$ where $0.28 <= x <= 0.4$) deposited at room temperature using DC magnetron co-sputtering. In close agreement with previous theoretical predictions, the SHR of the doped films was ~9% larger than the pure β-W films. We also found that the SHR's in stacks with $Co_2Fe_6B_2$ were nearly twice as large as the SHR's in stacks with $Co_4Fe_4B_2$ which we attribute to spin hall material and magnetic layer interface transparency. By optimizing the deposition parameters and using different substrates, the ideal $W_3Ta$ composition should be achievable and consequently attain the maximum SHR increase up to 20% relative to pure β-W films, paving the way to highly-efficient, commercially viable Pt-free, spintronic devices.

References

[1] Brataas A, Kent A D and Ohno H 2012 Current-induced torques in magnetic materials *Nat. Mater.* **11** 372–81

[2] Datta S, Salahuddin S and Behin-Aein B 2012 Non-volatile spin switch for Boolean and non-Boolean logic *Appl. Phys. Lett.* **101**

[3] Liu L, Moriyama T, Ralph D C and Buhrman R A 2011 Spin-torque ferromagnetic resonance induced by the spin Hall effect *Phys. Rev. Lett.* **106** 1–4

[4] Liu L, Pai C F, Li Y, Tseng H W, Ralph D C and Buhrman R A 2012 Spin-torque switching with the giant spin hall effect of tantalum *Science (80-. ).* **336** 555–8

[5] Pai C F, Liu L, Li Y, Tseng H W, Ralph D C and Buhrman R A 2012 Spin transfer torque devices utilizing the giant spin Hall effect of tungsten *Appl. Phys. Lett.* **101** 1–5

[6] Demasius K U, Phung T, Zhang W, Hughes B P, Yang S H, Kellock A, Han W, Pushp A and Parkin S S P 2016 Enhanced spin-orbit torques by oxygen incorporation in tungsten films *Nat. Commun.* **7** 1–7

[7] Hao Q and Xiao G 2015 Giant Spin Hall Effect and Switching Induced by Spin-Transfer Torque in a W/Co40Fe40 B20/MgO Structure with Perpendicular Magnetic Anisotropy *Phys. Rev. Appl.* **3** 1–6

[8] Zhu L, Ralph D C and Buhrman R A 2018 Highly Efficient Spin-Current Generation by the Spin Hall Effect in Au1-xPtx *Phys. Rev. Appl.* **10**

[9] Hägg G and Schönberg N 1954 `β-Tungsten' as a tungsten oxide *Acta Crystallogr.* **7** 351–2

[10] Morcom W R, Worrell W L, Sell H G and Kaplan H I 1974 The preparation and characterization of beta-tungsten, a metastable tungsten phase *Metall. Trans.* **5** 155–61

[11] Liu J and Barmak K 2016 Topologically close-packed phases: Deposition and formation mechanism of metastable β-W in thin films *Acta Mater.* **104** 223–7

[12] Barmak K and Liu J 2017 Impact of deposition rate, underlayers, and substrates on β-tungsten formation in sputter deposited films *J. Vac. Sci. Technol. A Vacuum, Surfaces, Film.* **35** 061516

[13] Gradhand M, Fedorov D V., Zahn P and Mertig I 2010 Extrinsic spin hall effect from first principles *Phys. Rev. Lett.* **104** 2–5

[14] Hoffmann A 2013 Spin hall effects in metals *IEEE Trans. Magn.* **49** 5172–93

[15] Sinova J, Valenzuela S O, Wunderlich J, Back C H and Jungwirth T 2015 Spin Hall effects *Rev. Mod. Phys.* **87** 1213–60

[16] Derunova E, Sun Y, Felser C, Parkin S S P, Yan B and Ali M N 2019 Giant intrinsic spin Hall effect in W3Ta and other A15 superconductors *Sci. Adv.* **5** 1–8

[17] Sui X, Wang C, Kim J, Wang J, Rhim S H, Duan W and Kioussis N 2017 Giant enhancement of the intrinsic spin Hall conductivity in β -tungsten via substitutional doping *Phys. Rev. B* **96** 1–5

[18] Zhang D, Wang Y and Gan Y 2013 Characterization of critically cleaned sapphire single-crystal substrates by atomic force microscopy, XPS and contact angle measurements *Appl. Surf. Sci.* **274** 405–17

[19] Kittel C 1948 On the theory of ferromagnetic resonance absorption *Phys. Rev.* **73** 155–61

[20]    Wang Y, Ramaswamy R and Yang H 2018 FMR-related phenomena in spintronic devices *J. Phys. D. Appl. Phys.* **51**

[21]    Nakayama H, Althammer M, Chen Y T, Uchida K, Kajiwara Y, Kikuchi D, Ohtani T, Geprägs S, Opel M, Takahashi S, Gross R, Bauer G E W, Goennenwein S T B and Saitoh E 2013 Spin Hall Magnetoresistance Induced by a Nonequilibrium Proximity Effect *Phys. Rev. Lett.* **110** 1–5

[22]    Zhang W, Han W, Jiang X, Yang S H and Parkin S S P 2015 Role of transparency of platinum-ferromagnet interfaces in determining the intrinsic magnitude of the spin Hall effect *Nat. Phys.* **11** 496–502

[23]    Pai C F, Ou Y, Vilela-Leão L H, Ralph D C and Buhrman R A 2015 Dependence of the efficiency of spin Hall torque on the transparency of Pt/ferromagnetic layer interfaces *Phys. Rev. B - Condens. Matter Mater. Phys.* **92**

[24]    Rojas-Sánchez J C, Reyren N, Laczkowski P, Savero W, Attané J P, Deranlot C, Jamet M, George J M, Vila L and Jaffrès H 2014 Spin pumping and inverse spin hall effect in platinum: The essential role of spin-memory loss at metallic interfaces *Phys. Rev. Lett.* **112** 1–5

[25]    Bass J and Pratt W P 2007 Spin-diffusion lengths in metals and alloys, and spin-flipping at metal/metal interfaces: An experimentalist's critical review *J. Phys. Condens. Matter* **19**

[26]    Wang Y, Deorani P, Qiu X, Kwon J H and Yang H 2014 Determination of intrinsic spin Hall angle in Pt *Appl. Phys. Lett.* **105**

[27]    Ikeda S, Miura K, Yamamoto H, Mizunuma K, Gan H D, Endo M, Kanai S, Hayakawa J, Matsukura F and Ohno H 2010 A perpendicular-anisotropy CoFeB-MgO magnetic tunnel junction *Nat. Mater.* **9** 721–4

[28]    Cuchet L, Rodmacq B, Auffret S, Sousa R C and Dieny B 2014 Influence of magnetic electrodes thicknesses on the transport properties of magnetic tunnel junctions with perpendicular anisotropy *Appl. Phys. Lett.* **105** 1–6

[29]    Belmeguenai M, Apalkov D, Gabor M, Zighem F, Feng G and Tang G 2018 Magnetic Anisotropy and Damping Constant in CoFeB/Ir and CoFeB/Ru Systems *IEEE Trans. Magn.* **54** 1–5

[30]    Lee C M, Ye L X, Chen H K and Wu T H 2013 The effects of deposition rate and annealing on CoFeB/MgO/CoFeB perpendicular magnetic tunnel junctions *IEEE Trans. Magn.* **49** 4429–32

[31]    Zhu M, Chong H, Vu Q B, Brooks R, Stamper H and Bennett S 2016 Study of CoFeB thickness and composition dependence in a modified CoFeB/MgO/CoFeB perpendicular magnetic tunnel junction *J. Appl. Phys.* **119**

**Supplement Figures:**

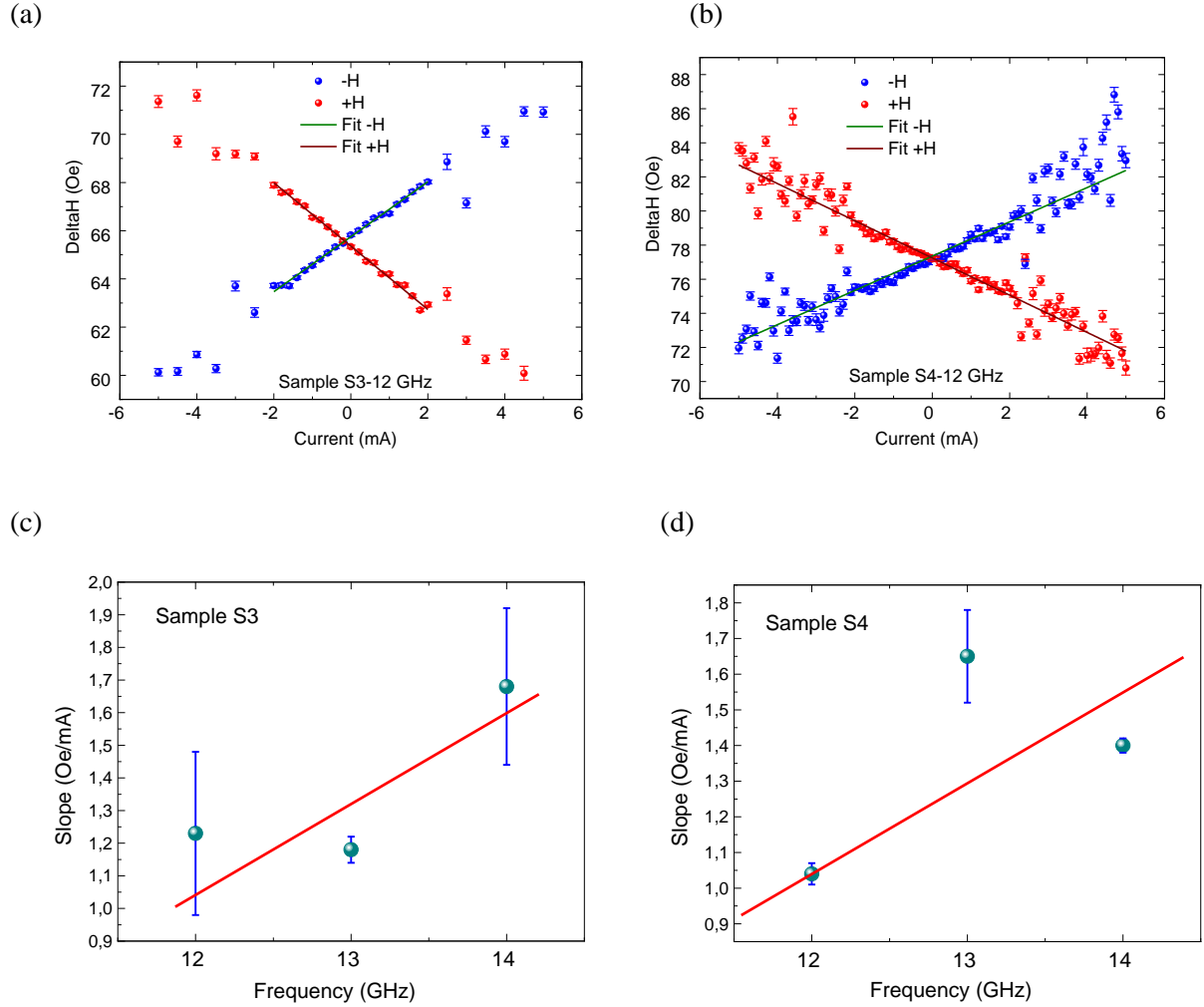

Figure 1:(a and b) The FMR linewidth as a function of DC current $I_{dc}$ for the W/Co$_4$Fe$_4$B$_2$ (S3) and W$_{4-x}$Ta$_x$/Co$_4$Fe$_4$B$_2$ (S4) bilayers device respectively at f=12 GHz. (c and d) Shows the slope as a function of frequency of the applied rf current in the bilayer device, the linear scaling is in accordance with equation (2).